\documentclass{elsart}

\usepackage{graphicx}
\usepackage{epsfig}

\usepackage{amssymb}
\begin{document}

\begin{frontmatter}

\title{Single-particle model for a granular ratchet}

\author[psu]{A. J. Bae}
\author[puc]{W. A. M. Morgado}\ead{welles@fis.puc-rio.br}
\author[psu]{J. J. P. Veerman}\ead{veerman@pdx.edu}
\author[ufpe]{G. L. Vasconcelos\corauthref{glv}}\ead{giovani@lftc.ufpe.br} 

\address[psu]{Mathematical Sciences, Portland State University,
Portland, OR 97207, USA.}
\address[puc]{Departamento de F\'{\i}sica, Pontif\'{\i}cia
Universidade Cat\'olica do Rio de Janeiro, CP 38071, 22452-970 Rio
de Janeiro, Brazil.}
\address[ufpe]{
Laborat\'orio de F\'{\i}sica Te\'orica e Computacional,
Departamento de F\'{\i}sica, Universidade Federal de Pernambuco,
50670-901, Recife, Brazil.}

\corauth[glv]{Corresponding author. Tel.: +55-81-3271-8450; 
fax: +55-81-3271-0359}

\begin{abstract}
A simple model for a granular ratchet corresponding to a single grain
bouncing off a vertically vibrating sawtooth-like base is
studied. Depending on the vibration strength, the sawtooth roughness
and the restitution coefficient, horizontal transport in both the
preferred and unfavoured directions is observed.  A phase diagram
indicating the regions in parameter space where each of the three
possible regimes (no current, normal current, and current reversal)
occurs is presented. 
\end{abstract}


\end{frontmatter}

\section{Introduction}

Ratchet-like motion, whereby directed transport is obtained from
nonequilibrium fluctuations, is important in many areas of scientific
and technological interest, such as, molecular motors in biology and
particle separation on nano- and microscales \cite{review}. Recently,
the problem of horizontal transport in a granular layer vertically
vibrated by an asymmetric base has also been considered, both
experimentally \cite{vicsek1} and computationally \cite{rapaport}. These
so-called ``granular ratchets'' have been shown to exhibit interesting
dynamical features akin to those found in Brownian-motor models.

In the present paper we approach the problem of granular ratchets by
considering the dynamics of a single grain bouncing off a vertically
vibrating sawtooth-shaped base. The model proposed here is a
generalization of a previous model studied by some of us
\cite{physicaD} for the gravity-driven motion of a particle on an
inclined rough surface with a staircase profile. In the present case,
the `rough surface' has a sawtooth shape with a horizontal baseline
and is subjected to a vertical vibration. We find that, in spite
of its simplicity, our single-particle model is able to reproduce some
of the generic behaviors observed in experiments \cite{vicsek1} and
computer simulations \cite{rapaport} of granular ratchets, such as
horizontal transport and current reversals.  The main results of the
paper are summarized in a phase diagram indicating the regions in
parameter space where the different dynamical regimes (no current,
normal current, and current reversal) can be observed.

\section{The model}
 
\begin{figure} 
\epsfxsize=8 cm
\begin{center}
\includegraphics*[width=0.8\columnwidth]{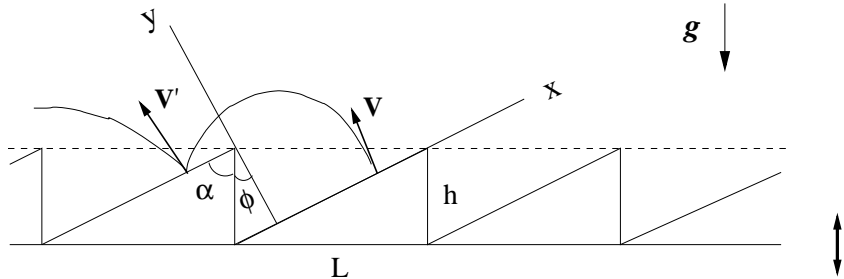}
\end{center}
\caption{A simple model for a granular ratchet.}
\label{fig1}
\vspace{0.3cm}
\end{figure}

Our granular ratchet model consists of a particle bouncing off a
vibrating sawtooth-like surface whose `teeth' have a vertical height
$h$ and a horizontal length $L$, as shown in Fig.~\ref{fig1}. A
particle is launched with given initial velocity over the vibrating
ratchet and then moves around through a succession of ballistic
flights and inelastic collisions.  We adopt a simple collision rule
given by $v_t'=v_t$ and $v_n'=-ev_n$, where $v_t$ and $v_n$ are
respectively the velocity components tangential and normal to the
collision plane, with prime denoting post-collisional velocities, and
$e$ is the coefficient of restitution.  We assume furthermore that the
sawtooth vibrates vertically with constant speed $V=2A/\tau$, where
$A$ is the vibration amplitude and $\tau$ its period.  For simplicity,
we consider that $A$ (and hence $\tau$) is negligibly small, so that the
sawtooth base remains essentially at the same height, which allows us
to determine explicitly the successive points of collision between the
particle and the ratchet.  The drawback here is that the effect of the
vibration on the particle velocity is  taken into account in a {\it
probabilistic} fashion, as will be discussed later. We believe,
however, that the dynamics seen in our model is quite robust and
should not be significantly altered if we consider a fully
deterministic model.

Suppose now that the particle takes off with velocity $\vec{v}=(u,v)$ from
a point $(x,y)$, which can be either on the sawtooth left edge
(inclined ramp) or on its right edge (vertical wall). For
convenience, we place our system of coordinates in such a way that the
$x$ axis is aligned with the inclined ramp and the $y$ axis makes an
angle $\phi=\frac{\pi}{2}-\alpha$, where $\alpha=\tan^{-1}(L/h)$, with
the vertical direction; see Fig.~\ref{fig1}.  (By convention, at each
new flight we translate our system of coordinates to the tooth from
which the particle departs.) Our  goal then is to determine the next
departure point $(x',y')$ and the corresponding takeoff velocity
$\vec{v}'=(u',v')$, so that the model dynamics can be described by a
map $$F:(x,y,u,v) \to (x',y',u',v').$$

To this end, let us introduce the following dimensionless
variables \cite{physicaD}
\[
\begin{array}{lll}
X= \displaystyle\frac{x}{b} \, , \quad & Y =
\displaystyle\frac{y}{a} \, ,
\quad &  \quad  T
= \displaystyle\sqrt{\frac{gc}{2a}} \, t \, , \\
U = \displaystyle \frac{c}{s} \displaystyle\frac{u}{\sqrt{2gca}}
\, , \quad & V = \displaystyle\frac{v}{\sqrt{2gca}} &  .  
\end{array}
\]
and the dimensionless roughness parameter $\kappa\equiv s^2/c^2$,
where $a=L\cos\alpha$, $b=L\sin\alpha$, $c\equiv\cos\phi$, and
$s\equiv\sin\phi$. Hereafter  we drop the capital notation with the
understanding that we shall work solely with dimensionless quantities.

Let us now define the jump number $n \in \mathbb{Z}$ in such a way
that its modulus equals the number of teeth the particle skips during
the flight, with the convention that $n$ is positive if the particle
moves in the direction of the positive $x$ axis.  To calculate $n$, we
must first determine the points (if any) where the particle trajectory
crosses the line $x+y=1$, connecting the tips of the sawteeth (the dashed line in
Fig.~1). Let $y_h^\pm$ be the $y$ coordinates of these two crossing
points (if they exist). An easy calculation shows \cite{us} that 
\[ y_h^\pm =
{1-x+\kappa y \over 1+\kappa} - 2\kappa(u-v) \frac{\kappa u+ v \pm
\sqrt{ (\kappa u+v)^2 - (1-x-y)(1+\kappa) }}{(1+\kappa)^2}.
\label{eq=y+-}
\] 
If $y_h^- \not \in (0,1]$ (including the case where there is no
real root) then the particle does not cross the line $x+y=1$ and hence
$n=0$. On the other hand, if $y_h^- \in (0,1]$ then the jump number is
determined by the second crossing point $y_h^+ $. More precisely, $n =
\left \lceil -y_h^+ \right \rceil$, where $\lceil x \rceil$ denotes
the ceiling function (i.e., the smallest integer greater than $x$).
Having thus determined $n$, we then need to find the point at which
the particle hits the surface. Note that if the particle jumps to the
right then it can land only on a ramp, but if the jump is to the left
then the particle can hit either a ramp or a vertical wall. Let
us first consider the case when the particle lands on a ramp.

\noindent {\it Collision with a ramp.}
Let $(x^*_r,y^*_r)$ denote the collision point and $(u^*_r,v^*_r)$ the
particle velocity prior to the collision. Using the kinematics of
ballistic motion and the fact that $y^*_r=-n$, one obtains
after a simple calculation that
\begin{equation}
\begin{array}{rcl}
u^*_r &=& u-v + \sqrt{v^2+n+y}, \\ v^*_r &=&
- \sqrt{v^2+n+y},\\
x^*_r &=& x - \kappa(n+y)+2\kappa(u-v)\left(v+\sqrt{v^2+n+y}\right) .
\end{array} \label{eq:before}
\end{equation}
If the particle jumps to the left (this happens when $u-v<0$), the
condition $n-x^*_r\le\kappa$ must also be satisfied, otherwise the
particle would have hit a vertical wall (this case will be treated
below).  On the assumption  that the particle lands on the $n$-th
ramp, the new departure point $(x',y')$ and the new tangential
velocity $u'$ after the collision will then be
\begin{equation}
x'=x^*_r-n,\quad y'=0,\quad u'=u^*_r.
\label{eq:afterI}
\end{equation}
Let us now calculate the normal velocity component $v'$ after the
collision.  First note that if $|v^*|\le\tilde{V}\equiv cV$ (here we
have dropped the subscript from $v^*_r$), then the particle can never
overtake the vibrating ramp and hence the collision will surely happen
when the sawtooth is moving upwards.  On the other hand, if
$|v^*|>\tilde{V}$ then the particle can hit the ramp either in its
upward or downward motion.  Furthermore, if the collision takes place
during the downward motion, then depending on the particle velocity
after the first collision there might  (or not) be enough time for the
sawtooth to revert its motion and hit the particle a second time.
Under the assumption that the vibration amplitude is negligibly small,
one can calculate the probabilities for all possible cases.  Here we
shall simply quote the final result and refer the interested reader to
Ref.~\cite{wamm_erm} for details of the calculation. If we define the
parameters $r=|v^*|/\tilde{V}$, $P_1=(1+r^{-1})/2$, and
$P_2=1-(1+e^{-1})/r$, then the outgoing velocity $v'$ and the
corresponding probability $p$ are as follows:
\begin{enumerate}
\item[i)]
${\rm if} \quad  r \in \left(0,1\right]  \quad {\rm
then}\quad v'= -ev^* + (1+e)\tilde{V}, \quad  p=1;$
\item[ii)]
${\rm if} \quad  r \in \left(1,1 + e^{-1}\right]  \quad {\rm
then}$
\begin{eqnarray*}
v'= -ev^* + (1+e)\tilde{V}, \quad && p=P_1,  \\
v'= e^2v^*+(1+e)^2\tilde{V}, \quad &&  p=1-P_1;
\label{eq=case3}
\end{eqnarray*}

\item[iii)]
${\rm if} \quad r\in (1+e^{-1},1+2e^{-1})\quad {\rm then}$
\begin{eqnarray*}
v'= -ev^* + (1+e)\tilde{V}, \quad &&  p=P_1,
\\ v'= -ev^* - (1+e)\tilde{V}, \quad && 
 p=P_2, \\
v'= e^2v^*+(1+e)^2\tilde{V}, \quad &&  p= 1-P_1-P_2;
\label{eq=case2}
\end{eqnarray*}

\item[iv)]
$ {\rm if} \quad r\ge 1 + 2e^{-1}   \quad {\rm then}$
\begin{eqnarray*}
 v'= -ev^* + (1+e)\tilde{V},\quad  &&   p=P_1,\\
v'= -ev^* - (1+e)\tilde{V},\quad  &&   p=1-P_1.
\label{eq=case1}
\end{eqnarray*}
\end{enumerate}
This probabilistic collision rule together with (\ref{eq:afterI}) thus
yield the map $F$ for the case when the collision occurs with a ramp.
Next we consider the case when the particle hits a vertical wall
(i.e., $u-v<0$ and $n-x^*_r>\kappa$).

\medskip
\noindent {\it Collision with a vertical wall.}
Here we first need to determine the actual collision point
$(x^*_w,y^*_w)$ with the vertical wall and the particle velocity
$(u^*_w,v^*_w)$ just before such collision. It is easy to show
\cite{us} that
\begin{equation}
\begin{array}{rcl}
x^*_w&=& x_0 + 2 \kappa  u t^* - \kappa t^{*2} ,  \qquad y^*_w= 2 v t^* - t^{*2} ,\\
u^*_w&=& u - t^* , \qquad v^*_w= v - t^* ,  
\end{array}
\label{eq=verticallanding}
\end{equation}
where
\begin{equation}
t^* = {\kappa(n-1)+n-x_0\over 2\kappa(u-v)} \quad .
 \label{eq=vertical}
\end{equation}

According to our convention,  the new departure point $(x',y')$ is given by
\begin{eqnarray}
x'&=&x^*_w-n,\label{eq:xyVert}
\\ y'&=&y^*_w+n.
\end{eqnarray}
Now, to determine the velocity $(u',v')$ after the collision we simply
apply the collision rule, $v_t'= v_t$ and $v_n'=-ev_n$, since the
vibration has no effect on the outgoing velocity in this case.  To do
that, we must first obtain $(v_t,v_n)$ in terms of $(u^*_w,v^*_w)$,
apply the above collision rule, and then return to the $(u',v')$
coordinates. Performing this calculation (details omitted), we find
\begin{eqnarray}
 u'&=& (s^2-ec^2) u^*_w+c^2(1 + e)v^*_w, \\ 
v' &=& s^2(1+e)u^*_w + (c^2-es^2)v^*_w.\label{eq:uvVert}
\end{eqnarray}
Equations (\ref{eq:xyVert})--(\ref{eq:uvVert}) give the map $F$ when
the collision takes place with a vertical wall, thus completing the
mathematical formulation of our model.  

\begin{figure}
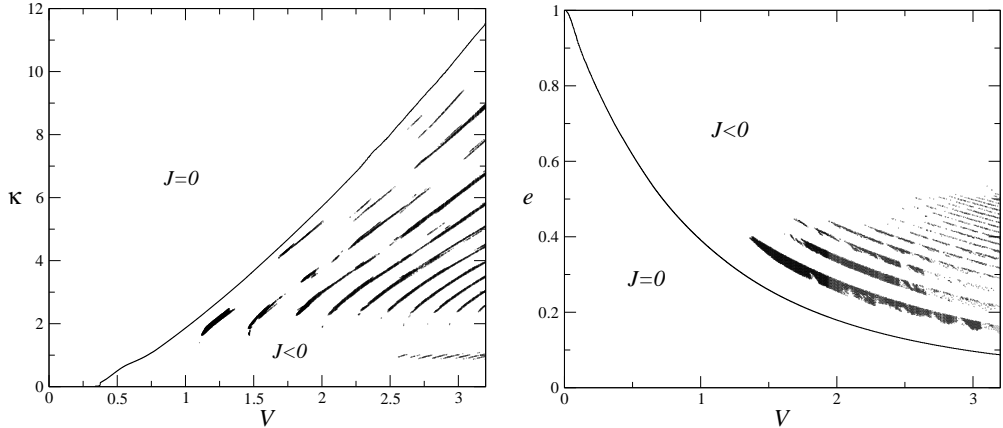
 
\epsfxsize=8 cm
\begin{center}
\includegraphics*[width=0.46\columnwidth]{fig2a.eps}\hspace{0.3cm}
\includegraphics*[width=0.46\columnwidth]{fig2b.eps}
\end{center}
\caption{Two-dimensional phase diagrams for the planes: (a) 
$(V,\kappa,e=0.5$) and (b) $(V,\kappa=1,e)$. The regions with $J=0$
and $J<0$ are separated by a critical line, and the dark regions
correspond to points where current reversal ($J>0$) occurs.}
\label{fig:diagram}
\end{figure}

\section{Phase diagram}

Although the map $F$ found above cannot be studied analytically owing
to the presence of a stochastic term, its long-term dynamics can be
easily investigated on the computer.  To this end, let us define the
current $J$ in our model as the particle average horizontal velocity:
$J\equiv\left<v_x\right>$, where the statistics is taken over a large
number of collisions. In the present paper, we are mainly interested in
mapping out the parameter space $(V,\kappa,e)$ according to which
dynamical regime ($J=0$, $J<0$, or $J>0$) is observed at a given
point. For illustration purposes, it is best however to look at 2D
slices of the parameter space and construct the corresponding phase
diagram.  Two such phase diagrams are shown in Figs.~2  for the
planes $(V,\kappa,e=0.5)$ and $(V,\kappa=1,e)$.

With the help of Fig.~2, the generic behavior of our ratchet model can
be summarized, as follows.  For given values of $\kappa$ and $e$, we
see that for sufficiently small $V$ there is no net current, i.e.,
$J=0$, meaning that the particle remains confined in one of the
``potential wells'' of the sawtooth profile. Then as $V$ increases
past a critical value $V_c(\kappa,e)$, we observe a transition to a
regime with $J<0$, where the particle, on average, moves horizontally
in the direction favoured by the ratchet asymmetry, namely, to the
left. In the two-dimensional phase diagrams of Fig.~\ref{fig:diagram},
the critical surface $V_c(\kappa,e)$ appears as a critical line
separating the region with $J=0$ from that with $J<0$. Furthermore,
for certain values of $V$, $\kappa$ and $e$, corresponding to the dark
regions in Fig.~2, a horizontal flow in the
`unfavoured' direction (i.e., with $J>0$) is established.  Notice that
these current-reversal regions appear as band-like structures,
a fact that can be traced back to the periodic nature of the ratchet
profile \cite{us}. It is important to note, however, that current
reversals can occur only for a sufficiently large $\kappa$ and
sufficiently small $e$. This is clearly seen in
Fig.~\ref{fig:diagram}, where there is no current reversal for
$\kappa<0.8$ in Fig.~2a nor for $e>0.6$ in Fig.~2b.

\section{Conclusions}

We have studied a simple model for a granular ratchet corresponding to
a single grain bouncing off a vertically vibrating sawtooth-like base.
We observe that for small shaking velocity $V$, the particle remains
confined between two teeth, with no net motion, whereas for larger $V$
a nonzero horizontal current is established. In this case, for most
values of the parameters the particle moves in the preferred
direction, but current reversals also occur rather regularly.  This
complex behavior is summarized in the phase diagram presented in
Fig.~2, which shows the regions in the parameter space $(V,\kappa,e)$
where each of the three possible regimes (no current, normal current,
and current reversal) can occur.  Our single-particle model is thus
able to reproduce some of the characteristic behavior observed in
experiments \cite{vicsek1} and computer simulations
\cite{rapaport} of  granular ratchets, such as horizontal
transport and current reversals.  In particular, our model shows that
for current reversal to occur two key ingredients are necessary: i) a
sufficiently rough vibrating surface and ii) sufficiently large
dissipation at collisions. Further investigation concerning the nature
of the `phase transition' to the nonzero-current regime as well as a
more detailed study of the current-reversal mechanism will be
presented in a forthcoming publication.

\begin{ack}
Financial support from the Brazilian agencies CNPq and FINEP and from
the special research program PRONEX is acknowledged.
\end{ack}


\begin{thebibliography}{399}

\bibitem{review} For a recent review, see, e.g., P.~Reimann, 
Phys.~Rep.~361 (2002) 57.


\bibitem{vicsek1} Z.~Farkas, P.~Tegzes, A.~Vukics, T.~Vicsek, 
Phys.~Rev.~E 60 (1999) 7022.

\bibitem{rapaport} M.~Levanon, D.~C.~Rapaport, Phys.~Rev.~E
64 (2001) 011304-1.

\bibitem{physicaD}
J.~J.~P.~Veerman, F.~V.~Cunha, Jr, G.~L.~Vasconcelos, Physica D
168-169 (2002) 220.

\bibitem{us} A.~J.~Bae, W.~A.~M.~Morgado, J.~J.~P.~Veerman, 
G.~L.~Vasconcelos, in preparation.

\bibitem{wamm_erm} W.~A.~M.~Morgado, E.~R.~Mucciolo, Physica A 
311 (2002) 150.


\end{thebibliography}
\end{document}